# MBE-grown 232–270 nm deep-UV LEDs using monolayer thin binary GaN/AlN quantum heterostructures

S. M. Islam, Kevin Lee, Jai Verma, Vladimir Protasenko, Sergei Rouvimov, Shyam Bharadwaj, Huili (Grace) Xing, and Debdeep Jena

# MBE-grown 232–270 nm deep-UV LEDs using monolayer thin binary GaN/AlN quantum heterostructures

S. M. Islam,[1,a)] Kevin Lee,[1] Jai Verma,[3,b)] Vladimir Protasenko,[1] Sergei Rouvimov,[4] Shyam Bharadwaj,[1] Huili (Grace) Xing,[1,2] and Debdeep Jena[1,2]

[1]*Department of Electrical and Computer Engineering, Cornell University, Ithaca, New York 14853, USA*
[2]*Department of Materials Science and Engineering, Cornell University, Ithaca, New York 14853, USA*
[3]*Department of Electrical Engineering, University of Notre Dame, Notre Dame, Indiana 46556, USA*
[4]*Notre Dame Integrated Imaging Facility, University of Notre Dame, Notre Dame, Indiana 46556, USA*

Electrically injected deep ultra-violet emission is obtained using monolayer thin GaN/AlN quantum structures as active regions. The emission wavelength is tuned by controlling the thickness of ultrathin GaN layers with monolayer precision using plasma assisted molecular beam epitaxy. Single peaked emission spectra are achieved with narrow full width at half maximum for three different light emitting diodes operating at 232 nm, 246 nm, and 270 nm. 232 nm (5.34 eV) is the shortest electroluminescence (EL) emission wavelength reported so far using GaN as the light emitting material and employing polarization-induced doping. *Published by AIP Publishing.* [http://dx.doi.org/10.1063/1.4975068]

Deep UV LEDs emitting photons of wavelength less than 280 nm are in strong demand. They have applications in several areas including water purification, medical diagnostics, and security. Replacing mercury based deep UV lamps by semiconductor LEDs is environmentally friendly and enables miniaturization, higher energy efficiency and portability, faster turn-on, and integration. The family of high Al composition $Al_xGa_{1-x}N$ alloy (x > 0.5) semiconductors is well suited for such LEDs. Deep UV emission from LEDs using the AlGaN multi-quantum well (QW) active light emitting region has been obtained in several reports.[1–7] Although AlGaN UV-C LEDs spanning the 210–280 nm spectral window have been demonstrated, they suffer from low external quantum efficiency (EQE),[8] which drops precipitously for wavelengths shorter than 240 nm.

The reasons for this efficiency drop are manifold. The LEDs are typically grown on non-native substrates such as sapphire with a large lattice mismatch, leading to dislocation densities $10^9 cm^{-2}$. Non-radiative recombination in the active quantum well regions causes a dramatic decrease in the internal quantum efficiency (IQE). Spontaneous and piezoelectric polarization-induced internal electric fields in the active region quantum well heterostructures spatially separate the electron and hole wavefunctions, reducing the overlap integral and the oscillator strength for interband radiative transitions.[9–11] This quantum-confined Stark effect (QCSE) reduces the IQE, adding to non-radiative recombination at threading dislocations.

Electrical injection of electrons and holes into the active region is challenging in high Al-composition AlGaN due to the high activation energy of common dopants. The low thermal activation of free carriers at room temperature, especially for doping using Mg, causes highly resistive carrier injection layers, limiting the wall-plug efficiency due to high voltage drops.[4–6] As the Al content increases in the $Al_xGa_{1-x}N$ alloys, the polarization of emitted light changes from surface-emitting TE (electric field of the optical wave is normal to the c-axis) to edge-emitting TM (electric field parallel to the c-axis).[12] Surface-light extraction becomes challenging for high Al content AlGaN alloy active regions that are necessary for sub 260 nm emission.

As an alternative to AlGaN, the use of ultra-thin GaN quantum wells and dots as the light emitting material has been proposed.[13–15] By thinning down from bulk GaN of bandgap 365 nm (3.4 eV) to a few monolayer (ML) thick quantum structures sandwiched between AlN barriers, tunable deep UV photon emission from 234–274 nm (~5.3–4.5 eV) was observed.[15] This extreme blue-shift due to quantum confinement is possible because of the large band offset between AlN and GaN. The emission wavelength can be further lowered to 224 nm (~5.53 eV) for 1 ML GaN quantum wells[16,17] and further down to 222 nm (~5.6 eV) with 1–2 ML quantum dots (QDs).[18] Such ultrathin quantized active regions can improve the IQE.[18,19] The QCSE is suppressed because a negligible voltage may drop across a thin layer; the overlap integral is maximized. Defect and composition fluctuation-free fully strained monolayer-thick GaN quantum structures improve the IQE over an AlGaN alloy active region. Because of three-dimensional confinement of carriers in quantum dots, injected carriers are kept away from non-radiative recombination sites at dislocations, further enhancing the IQE.[20] Finally, the emitted light from the ML thin GaN quantized structures is surface-emitting strongly TE polarized,[21] improving light extraction compared to AlGaN active regions. These GaN deep-UV active regions also offer integration pathways to overcome carrier injection bottlenecks for LED devices. The use of internal polarization to field-ionize or activate carriers from deep dopants is achieved by using either short period superlattices[21,22] or polarization-induced doping[23] in graded AlGaN injection layers.

In this letter, we demonstrate electrically injected tunable deep UV LEDs using ultra-thin GaN quantum dots in the active region. By controlling the GaN layer thicknesses

a)Author to whom correspondence should be addressed. Electronic mail: smislam@cornell.edu
b)Present address: Intel Corporation, Oregon, USA.

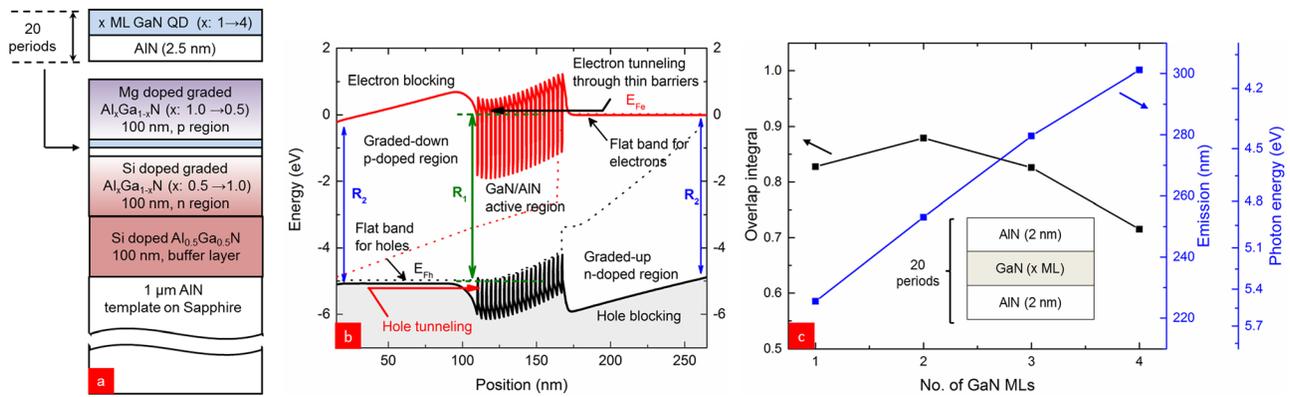

FIG. 1. (a) Heterostructure layer design of the deep-UV LED, (b) the simulated energy band diagram at 5 V forward bias, and (c) Calculated evolution of UV emission wavelength and the wavefunction overlap integral with GaN ML thickness.

with ML precision and employing a polarization-induced doping scheme for both n and p carrier injection regions to enhance the electrical injection of carriers into the active region,[23] the shortest electroluminescence (EL) emission wavelength achieved in this work is 232 nm (~5.34 eV), compared to the last reported 243 nm (~5.1 eV).[15] This makes it the record short wavelength UV LED in a structure that combines the binary GaN active region with polarization-induced doping.

The LED quantum heterostructures were grown in a Veeco Gen930 RF plasma molecular beam epitaxy (MBE) system on commercial 1 $\mu$m thick metal-polar AlN templates on 430 $\mu$m thick sapphire, which had a typical threading dislocation density ~$10^{10}$/cm$^2$. The templates were In-mounted on unpolished 3-in. Si carrier wafers and cleaned in organic solvents prior to loading into the MBE system. A 200 °C/8 h bake in the load-lock chamber followed by a 450 °C/90 min bake in the buffer chamber was performed to desorb atmospheric contaminants prior to transfer into the growth chamber, where they were heated to 700 °C as read by a thermocouple to initiate epitaxial growth. Fig. 1(a) shows the epitaxial layer structure that was grown for the UV LEDs reported in this work. A 100 nm thick Si-doped 50% AlGaN n-contact layer was grown on the template, followed by a 100 nm Si doped linearly graded AlGaN electron injection layer. The Si doping was ~$10^{20}$/cm$^3$, and the Al composition was graded from 50% to 100% to take advantage of polarization-induced doping. Separate calibration growths combined with X-Ray diffraction were performed to obtain the desired Al composition and grading. For the graded AlGaN layers, both the Al and Ga fluxes were adjusted simultaneously to keep the surface free from excessive Ga accumulation as tracked by in-situ reflection high-energy electron diffraction (RHEED).

After the n-type graded layer, 20 periods of GaN/AlN active regions were grown at a substrate temperature of 700 °C. The growth rate was ~0.3 ML/s throughout, at a 400 W Nitrogen plasma power. The GaN quantum dots were grown using the Stranski-Krastanov growth method as confirmed by RHEED.[24,25] To ensure 3D GaN island formation, a N-rich active flux ratio Ga/N < 1 was maintained in the active region. The thickness of the GaN quantum dots was varied from 1–4 MLs by varying the growth time and the Ga/N ratio[15] over 3 different samples [D1 (1–2 MLs), D2 (2–3 MLs), and D3 (3–4 MLs)]. The AlN barrier thickness was kept at 2 nm in each period to simultaneously provide sufficient quantum confinement, while allowing tunneling transport between dots. These AlN layers were grown using migration enhanced epitaxy (MEE) by first depositing 8 MLs of Al in the absence of active nitrogen flux, and then supplying Nitrogen flux in the absence of Al flux. The consumption of all excess Al was confirmed by observing characteristic transitions in the RHEED streaks. The MEE technique enabled the growth of smooth 2D AlN layers at a relatively low growth temperature of 700 °C. The active region was then capped with a 100 nm $10^{19}$/cm$^3$ Mg-doped compositionally graded p-AlGaN hole injection layer graded down from 100% to 50%, and grown at 630 °C for sufficient Mg incorporation.

The simulated energy band diagram of the LED structure using SiLENSe at a forward bias of 5 V is shown in Fig. 1(b). The simulation estimates the expected radiative recombination energies in the all-binary AlN/GaN active region $R_1$, as well as in the n- and p-carrier injection regions indicated by $R_2$, to be compared to experiment. The electron and hole quasi-Fermi levels $E_{fe}$ and $E_{fh}$ and their separation indicate the location of radiative recombination, and the bands indicate the built-in electron and hole blocking layers because of grading. The flat bands in the carrier injection layers ensure smooth carrier transport up to the active region, followed by tunnel injection. The simulated evolution of the photon emission wavelength and corresponding overlap integral for the GaN quantum well (QW)/AlN heterostructure is shown in Fig. 1(c). The simulation indicates that as the GaN thickness is reduced from 4 ML → 3 ML → 2 ML → 1 ML, the emission wavelength evolves as 301 nm → 280 nm → 253 nm → 225 nm. The overlap integral also approaches a maximum at 88% for 1–2 ML thick GaN quantum wells in AlN barriers, which suggests that the IQE for such thin GaN layers should improve significantly compared to thicker GaN (or AlGaN) wells due to the QCSE. A maximum in the IQE is to be expected from heuristic grounds in polar quantum wells, because as the well thickness increases, the QCSE decreases the IQE whereas reducing the well thickness causes carrier wavefunction penetration into the barrier and leakage. The simulation indicates that for the AlN/GaN/AlN structure, the optimal optical recombination is to be expected for a well thickness of ∼2 MLs. Although this well thickness fixes the emission wavelength, further blue-shift may be achieved by forming dots and taking advantage of in-plane quantum confinement.

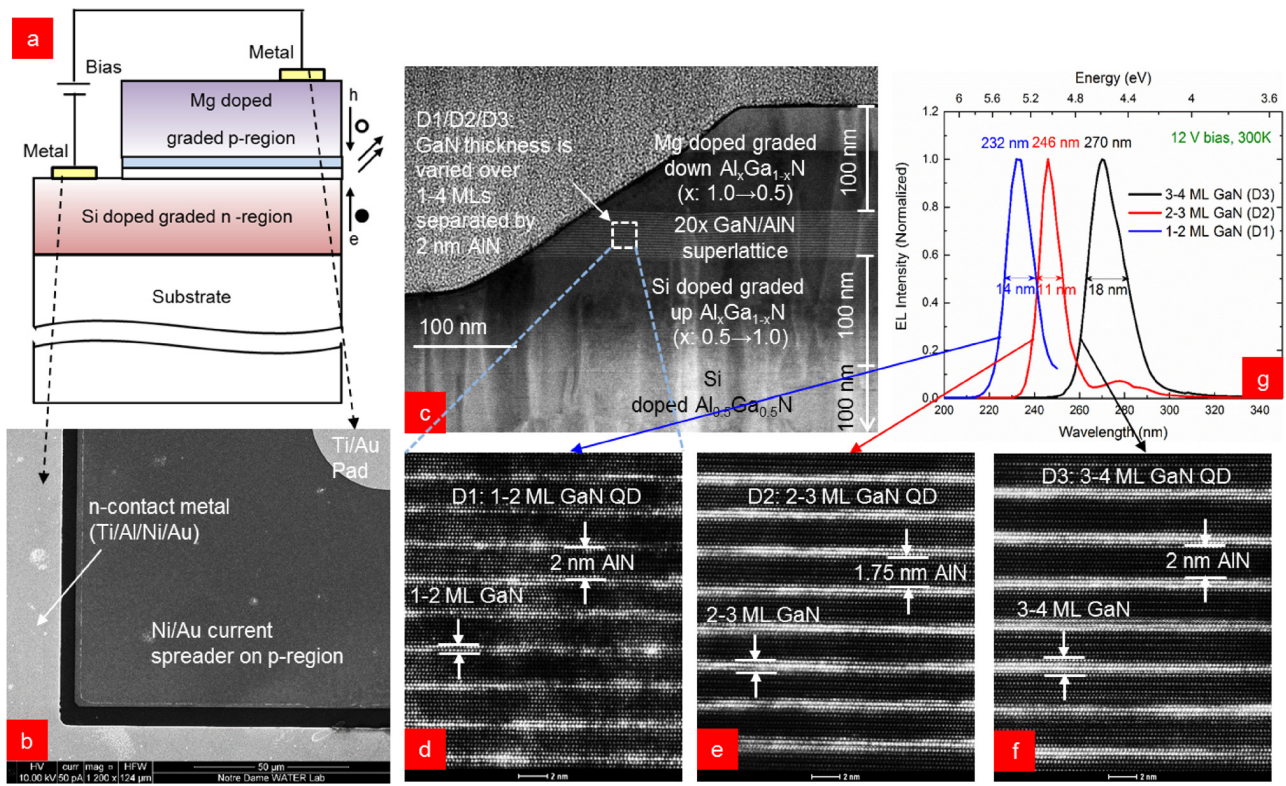

FIG. 2. (a) Schematic of the processed device and (b) SEM (top view) of a processed device showing n and p metal contacts. This is one quadrant view of the whole device, (c) cross section HAADF-STEM image of a processed device showing different layers for a typical sample (D1/D2/D3), zoomed in active region for (d) D1, (e) D2, (f) D3, and (g) EL spectra taken at 12 V forward bias for D1–D3 at 300 K.

Deep-UV LEDs were fabricated by optical lithography followed by mesa etching 200 nm in $Cl_2$ based reactive-ion etch and e-beam evaporation of a Ti (20 nm)/Al (100 nm)/Ni (40 nm)/Au (50 nm) metal stack on the n-AlGaN layer and a Ni (5 nm)/Au (5 nm) current spreading layer followed by a 50 μm diameter circular Ti (20 nm)/Au (100 nm) probe pad for the p-type layer as indicated in Fig. 2(a). 200 × 200 μm LEDs shown in the SEM image in Fig. 2(b) were realized. Fig. 2(c) shows the entire cross-section TEM image with all layers of the UV LED device D2 which had 2–3 ML GaN quantum dots in the active region, while Fig. 2(d) shows the high-resolution atomic image of the 2 nm AlN barriers and 1–2 ML GaN quantum dot active regions. In addition to the active region control, the TEM images prove the epitaxial growth of the graded n- and p-AlGaN layers, and Figs. 2(d), 2(e), and 2(f) prove that ultrathin GaN quantum heterostructures of the desired thicknesses 1–2 MLs (D1), 2–3 MLs (D2), and 3–4 MLs (D3) were achieved by MBE. Fig. 2(g) shows the measured room-temperature electroluminescence spectra for the three LEDs under pulsed excitation at 10 kHz, 5% duty cycle. All LEDs showed strong single peak emission. A blue shift in the EL peak from 270 nm → 246 nm → 232 nm is observed as the thickness of the active region GaN layer is reduced.

Fig. 3(a) shows a comparison of the measured EL spectra with the simulated EL spectra for 1 → 2 → 3 ML GaN quantum well LEDs, indicating qualitative agreement in the wavelengths. The discrepancies in wavelength and the lineshape of the emission are because of the inhomogeneous quantum dot distribution and quantum well and barrier thickness variation in experiment and limitations of the band structure at high conduction and valence band energies in the theory. The light emitting active regions of the LEDs consisted of 1–2 MLs, 2–3 MLs, and 3–4 MLs of GaN for the 3 samples, respectively, meaning that there should exist thickness-variation dependent emission broadening for each of these devices. Thicker quantum well regions have lower quantum confinement and thus red-shifted photon emission. The light emission process is governed by the thickest part of the GaN region because of the fact that the thermally assisted diffusive transport of electrons and holes at room temperature make them accumulate at the smallest energy regions before they can radiatively recombine. Therefore, the extended tail of the spectra in the longer wavelength side is determined by the thicker regions of the quantum wells. Also this effect is prominent for the 1–2 ML sample because the rate of change of emission energy with GaN thickness (dE/dz) is the highest for this sample among the three. The shortest EL wavelength reported so far using the GaN active region is 239 nm by Metal-Organic Chemical Vapor Deposition (MOCVD)[14] and 243 nm by MBE.[15] This work demonstrates that by changing the thickness of the GaN layer with a ML precision, it is possible to achieve tunable deep UV LEDs emitting as short as 232 nm.

The EL data shown in Fig. 2(g) were measured with a 12 V forward bias and with a current density of 165 A/cm$^2$ (270 nm), 90 A/cm$^2$ (246 mA), and 290 A/cm$^2$ (232 nm) for the three devices. The measured current densities are composed of the leakage current through dislocations and the component of current that causes light emission. The turn-on voltages for the devices were ~7.5 V. Contact and sheet resistance for the n-AlGaN regions were characterized by the transmission-line-measurement method on separately grown calibration samples and values of $1.2 \times 10^{-4}$ $\Omega$ cm$^2$ (contact resistance) and $2.76 \times 10^{-3}$ $\Omega$ cm (sheet resistance) were obtained. The uniform current spreading was confirmed by observing the light emission pattern from the devices during the EL measurement as shown in the inset of Fig. 3(a). The reason for the higher-than-expected turn-on voltages are poor p-type contacts and the use of 20 periods in the active region. Since the light emitting active region consists of a few monolayers of GaN, in this experiment 20 periods have been used to increase the effective volume of the light emitting region. Based on some recent simulation and experimental work, 8 periods of the active region seem to be the optimum in terms of the light output and forward voltage drop across the device. Reducing the number of periods and further improvements in the p-layer design using tunnel-contacts[26,27] can potentially help improve the hole injection problem for deep-UV LEDs.

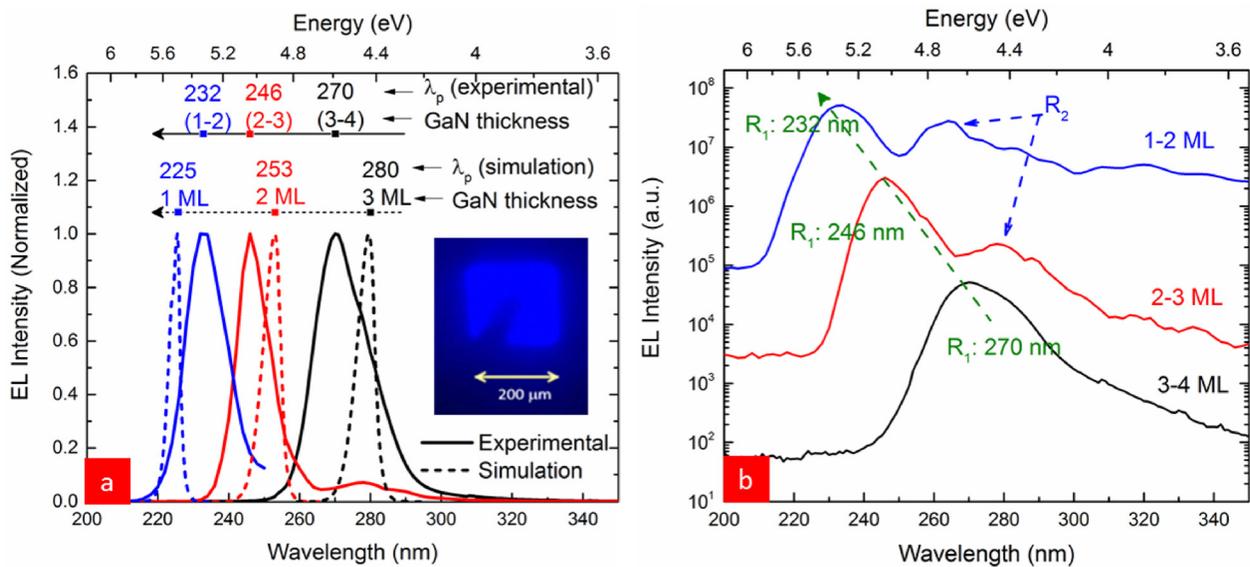

FIG. 3. (a) Comparison of measured and simulated EL spectra showing qualitative agreement in the shift of the wavelengths. The inset shows the uniform light distribution pattern from the device confirming good current spreading. (b) The measured EL spectra in log scale showing secondary $R_2$ peaks from re-absorption in the AlGaN carrier injection layers for D1 and D2.

Fig. 3(b) shows the logarithmic plot of the EL spectra for the three LEDs. For D1 with 1–2 ML quantum dot active regions and D2 with 2–3 ML quantum dots, a secondary EL peak $R_2$ is observed at ~4.6 eV in addition to the active region peak $R_1$. Fig. 1(b) shows the radiative processes leading to these peaks. The EL spectra were collected from the top of the devices. The graded n- and p-AlGaN regions have thin regions of 50% Al content AlGaN to boost the polarization-induced doping and conventional doping. The bandgap of Al$_{0.5}$Ga$_{0.5}$N is 4.6 eV,[28] which leads to the reabsorption of the 232 nm and 246 nm photons emitted from the active regions of D1 and D2 LEDs and re-emission after the downconversion. The secondary emission peak $R_2$ is relatively stronger for D1 than for D2 possibly because of the higher absorption cross-section at a higher photon energy of the $R_1$ peak (5.34 eV for D1 vs 5.04 eV for D2). However, because Al$_{0.5}$Ga$_{0.5}$N is transparent to 270 nm, LED D3 shows single $R_1$ peak EL spectra. The difference in the $R_2$ peak positions for D1 and D2 devices can be due to slight compositional mismatch of the graded layer growths. As outlined in Ref. 15, the transparency of the $R_1$ peak can be achieved for D1 and D2 devices as well by choosing the appropriate minimum composition of p-AlGaN, i.e., Al$_{0.7}$Ga$_{0.3}$N. This will ensure transparency but will compromise the boost expected by polarization-induced doping.

In summary, tunable deep UV electroluminescence over 232–270 nm (5.34–4.59 eV) is demonstrated using few ML thick GaN quantum structures with AlN barriers using MBE. Emission wavelength as short as 232 nm has been achieved using 1–2 ML GaN quantum structures with a spectral linewidth of 14 nm at 300 K. The plasma-MBE technique was used to maintain precise control on the thickness of the grown epitaxial layers. Tunneling transport through the barriers and polarization-induced doping technique have been used to enhance light emission from the devices. The alternative method to obtain deep-UV LEDs shown in this work potentially offers solutions to several problems that limit the quantum efficiencies of current AlGaN quantum well based devices, such as surface vs edge emission, elimination of the quantum confined Stark effect, significant robustness to the presence of defects, and the ability to engineer the wavelength of emission using all-binary heterostructures. The reduced dimensionality of the active region can also be suitable for amplified spontaneous emission for LEDs and perhaps lower the threshold for realizing electrically injected deep-UV Lasers, which remains an unsolved problem in semiconductor physics.

The authors acknowledge discussions with D. Bayerl and M. Kioupakis from University of Michigan. This work was supported in part by a NSF DMREF grant #1534303, and an AFOSR grant monitored by Dr. Ken Goretta.